\documentclass[conference]{IEEEtran}
\IEEEoverridecommandlockouts

\ifCLASSINFOpdf
\else
\fi

\usepackage[english]{babel}
\usepackage{times,amssymb,amsmath,amsfonts,float,color}
\usepackage{euscript,graphics}
\usepackage[dvips]{epsfig}

\usepackage{hyperref}
\usepackage{amsthm}
\usepackage{multirow}

\theoremstyle{plain}
\newtheorem{theorem}{Theorem}

\newtheorem{lemma}{Lemma}
\newtheorem{proposition}{Proposition}
\newtheorem{corollary}[proposition]{Corollary}
\newtheorem{definition}{Definition}

\newtheorem{cnstr}{Construction}

\newtheorem{xmpl}{Example}

\newcommand{\remove}[1]{}

\newcommand\nc\newcommand
\nc\bfa{{\boldsymbol a}}\nc\bfA{{\bf A}}\nc\cA{{\mathcal A}}
\nc\bfb{{\boldsymbol b}}\nc\bfB{{\bf B}}\nc\cB{{\mathcal B}}
\nc\bfc{{\boldsymbol c}}\nc\bfC{{\bf C}}\nc\cC{{\mathcal C}}
\nc\bfd{{\boldsymbol d}}\nc\bfD{{\bf D}}\nc\cD{{\mathcal D}}
\nc\bfe{{\boldsymbol e}}\nc\bfE{{\bf E}}\nc\cE{{\mathcal E}}
\nc\bff{{\boldsymbol f}}\nc\bfF{{\bf F}}\nc\cF{{\mathcal F}}
\nc\bfg{{\boldsymbol g}}\nc\bfG{{\bf G}}\nc\cG{{\mathcal G}}
\nc\bfh{{\boldsymbol h}}\nc\bfH{{\bf H}}\nc\cH{{\mathcal H}}
\nc\bfi{{\boldsymbol i}}\nc\bfI{{\bf I}}\nc\cI{{\mathcal I}}
\nc\bfj{{\boldsymbol j}}\nc\bfJ{{\bf J}}\nc\cJ{{\mathcal J}}
\nc\bfk{{\boldsymbol k}}\nc\bfK{{\bf K}}\nc\cK{{\mathcal K}}
\nc\bfl{{\boldsymbol l}}\nc\bfL{{\bf L}}\nc\cL{{\mathcal L}}
\nc\bfm{{\boldsymbol m}}\nc\bfM{{\bf M}}\nc\cM{{\mathcal M}}
\nc\bfn{{\boldsymbol n}}\nc\bfN{{\bf N}}\nc\cN{{\mathcal N}}
\nc\bfo{{\boldsymbol o}}\nc\bfO{{\bf O}}\nc\cO{{\mathcal O}}
\nc\bfp{{\boldsymbol p}}\nc\bfP{{\bf P}}\nc\cP{{\mathcal P}}
\nc\bfq{{\boldsymbol q}}\nc\bfQ{{\bf Q}}\nc\cQ{{\mathcal Q}}
\nc\bfr{{\boldsymbol r}}\nc\bfR{{\bf R}}\nc\cR{{\mathcal R}}
\nc\bfs{{\boldsymbol s}}\nc\bfS{{\bf S}}\nc\cS{{\mathcal S}}
\nc\bft{{\boldsymbol t}}\nc\bfT{{\bf T}}\nc\cT{{\mathcal T}}
\nc\bfu{{\boldsymbol u}}\nc\bfU{{\bf U}}\nc\cU{{\mathcal U}}
\nc\bfv{{\boldsymbol v}}\nc\bfV{{\bf V}}\nc\cV{{\mathcal V}}
\nc\bfw{{\boldsymbol w}}\nc\bfW{{\bf W}}\nc\cW{{\mathcal W}}
\nc\bfx{{\boldsymbol x}}\nc\bfX{{\bf X}}\nc\cX{{\mathcal X}}
\nc\bfy{{\boldsymbol y}}\nc\bfY{{\bf Y}}\nc\cY{{\mathcal Y}}
\nc\bfz{{\boldsymbol z}}\nc\bfZ{{\bf Z}}\nc\cZ{{\mathcal Z}}
\nc\od{{\bar d}}\nc\ow{{\bar w}}\nc\odelta{{\bar\delta}}
\nc\ox{{\bar x}}\nc\oy{{\bar y}}\nc\ou{{\bar u}}
\nc\oh{{\bar h}}
\nc{\tr}{{T}}

\newcommand\ff{{\mathbb F}}

\nc\ellone{{\ell_1}}
\nc\elltwo{{\ell_2}}
\nc\ellinf{{{\ell_\infty}}}
\nc\ip[2]{\langle #1,#2\rangle}

\newcommand{\beeq}{\begin{eqnarray*}}
\newcommand{\eneq}{\end{eqnarray*}}

\begin{document}
%
\title{Locally Repairable Codes with Multiple \\
$(r_{i}, \delta_{i})$-Localities}

\author{\IEEEauthorblockN{Bin Chen \IEEEauthorrefmark{1}\IEEEauthorrefmark{2}\thanks{\IEEEauthorrefmark{1}The authors are with the Graduate School at Shenzhen, Tsinghua University, Shenzhen 518055, China. (emails: binchen14scnu@m.scnu.edu.cn, xiast@sz.tsinghua.edu.cn, j-hao13@mails.tsinghua.edu.cn.)
This work is supported in part by the National Natural Science Foundation of China under grant No. 61371078, and the R\&D Program of Shenzhen under grant Nos. JCYJ20140509172959977, JSGG20150512162853495, ZDSYS20140509172959989, JCYJ20160331184440545.
}\thanks{\IEEEauthorrefmark{2}Bin Chen is with the School of Mathematical Sciences, South China Normal University, Guangzhou, China. This research was done at the Graduate School at Shenzhen, Tsinghua University.}}
\and
\IEEEauthorblockN{Shu-Tao Xia\IEEEauthorrefmark{1}}
\and
\IEEEauthorblockN{Jie Hao\IEEEauthorrefmark{1}}}


%


\maketitle

\begin{abstract}
In distributed storage systems, locally repairable codes (LRCs)  are introduced  to realize low disk I/O and repair cost. In order to tolerate multiple node failures, the LRCs with \emph{$(r, \delta)$-locality} are further proposed. Since hot data is not uncommon in a distributed storage system, both Zeh \emph{et al.} and Kadhe \emph{et al.} focus on the LRCs with \emph{multiple localities or unequal localities} (ML-LRCs) recently, which said that the localities among the code symbols can be different. ML-LRCs are attractive and useful in reducing repair cost for hot data.
In this paper, we generalize the ML-LRCs to the $(r,\delta)$-locality case of multiple node failures, and define an LRC with multiple $(r_{i}, \delta_{i})_{i\in [s]}$ localities ($s\ge 2$), where $r_{1}\leq r_{2}\leq\dots\leq r_{s}$ and $\delta_{1}\geq\delta_{2}\geq\dots\geq\delta_{s}\geq2$. Such codes ensure that some hot data could be repaired more quickly and have better failure-tolerance  in certain cases because of relatively smaller $r_{i}$ and larger $\delta_{i}$.  Then, we derive a Singleton-like upper bound on the minimum distance for the proposed LRCs by employing the regenerating-set technique. Finally, we obtain a class of explicit and structured constructions of optimal ML-LRCs, and further extend them to the cases of multiple $(r_{i}, \delta)_{i\in [s]}$ localities.
\end{abstract}


%
\IEEEpeerreviewmaketitle

\section{Introduction}

Recently, locally repairable codes (LRCs) have attacted a lot of interest.  Let $\mathbb{F}_q$ be a finite field with size $q$. The $i${th} symbol $c_i$ of a $q$-ary $[n, k]$ linear code $\cC$ said to have \emph{locality} $r$ if this symbol can be recovered by accessing at most $r$ other symbols of $\cC$ \cite{gopalan2011locality}. In distributed storage systems, $r \ll k$ indicates that only a small number of storage nodes are involved in repairing a failed node, which means low disk I/O and repair cost. The code is called an $(n,k,r)$ LRC or an $r$-local LRC. It was shown that the minimum distance $d$ is upper bounded by \cite{gopalan2011locality}
\begin{equation}
\label{singleton}
d\leq n-k-\left\lceil{k}/{r}\right\rceil+2.
\end{equation}
An LRC meeting this Singleton-like bound is called \emph{optimal}. Various constructions of optimal LRCs  were obtained recently, e.g.,\cite{gopalan2011locality}-\cite{Tamo matroidIT}.

In order to tolerate multiple node failures in a distributed storage system, an important extension to the $r$-local codes is the so-called LRC with $(r, \delta)$-locality \cite{prakash2012optimal}. According to \cite{prakash2012optimal}, the $i$-{th} symbol $c_i$ of a $q$-ary $[n, k]$ linear code $\cC$ is said to have $(r, \delta)$-locality ($\delta\ge2$) if  there exists a punctured subcode of $\cC$ with support containing $i$, whose length is at most $r + \delta - 1$, and whose minimum distance is at least $\delta$, i.e., there exists a subset $S_{i}\subseteq [n]\triangleq\{1,2,\ldots,n\}$ such that $i\in S_{i}$, $|S_{i}|\le r+\delta-1$ and $d_{min}(\cC|_{S_{i}})\ge\delta$. The code $\cC$ is said to have $(r, \delta)$-locality or be a $(r,\delta)$-LRC if all the symbols have $(r,\delta)$-localities. And a Singleton-like bound was also obtained, which said that
\begin{equation} \label{eq_GeneralizedSingleton}
d \leq n-k+1-\left( \left \lceil {k}/{r} \right \rceil-1 \right)(\delta -1).
\end{equation}
The codes meeting it are called optimal $(r,\delta)$-LRCs. Note that it degenerates to (\ref{singleton}) when $\delta =2$. Many optimal constructions of $(r,\delta)$-LRCs can be founded in \cite{Tamo matroidIT}-\cite{generalized cyclic}.

Codes with \emph{multiple localities or unequal localities }  were firstly introduced in \cite{zeh} and \cite{kadhe}, which said that the locality among the code symbols can be different. Such an LRC with multiple localities is practically appealing in hot data (i.e., the data is accessed more frequently) that need to be repaired quickly and thus require smaller locality. More specifically, a code $\cC$ with unequal information locality \cite{kadhe} was interpreted as follows: the information locality profile of an $[n,k,d]$ linear code $\cC$ is defined by a length-$r$ vector $\textbf{k}(\cC) = \{k_{1}, \ldots, k_{r}\}$, where $k_{j}$ is the number of information symbols with locality $j$ for $j\in[r]$. Clearly,  $\forall j\in[r]$, $0\le k_j\le k$, $k_{r} \geq 1$ and $\sum_{j=1}^{r} k_{j} = k$. An upper bound on the minimum distance was obtained as
\begin{equation}
\label{eq:bound-1}
d \le n - k - \sum_{j=1}^{r}\left\lceil{{k_{j}}/{j}}\right\rceil + 2.
\end{equation}
The code with all-symbol multiple localities or unequal all-symbol locality was introduced respectively in \cite{zeh} and \cite{kadhe}. Two different forms of upper bounds on the minimum distance were also obtained according to different restrictive conditions. In this paper, we adopt the definition of the LRC with all-symbol multiple localities (ML-LRC) in \cite{zeh}. Let $s\geq 2$ and $\mathcal{T}_1, \mathcal{T}_2, \ldots, \mathcal{T}_s$ be a partition of $[n]$, i.e., $\cup_{i \in [s]} \mathcal{T}_{i}= [n]$ and $\mathcal{T}_i\cap \mathcal{T}_j=\emptyset$. Let $n_i=| \mathcal{T}_i|$. A $q$-ary $[n,k,d]$ linear code is called $((n_1, r_{1}), (n_2, r_{2}), \dots, (n_{s}, r_{s}))$-local ($r_{1}<r_{2}<\dots<r_{s}$) if each code symbols in a set $\mathcal{T}_{i}$ are a linear combination of at most $r_{i}$ other code symbols within $\mathcal{T}_{i}$ for all $i \in [s]$. In \cite{zeh}, a Singleton-like upper bound for the code is obtained, i.e., if $\sum_{i\in[s-1]}r_i\lceil n_{i}/(r_{i}+1)\rceil<k-1$, then
\begin{equation} \label{eq_SingletonBoundMultipleLocalities}
d  \le\! n - k + 2 -\!\!\!\!\sum_{i \in [s-1]} \!\!\left \lceil \frac{n_i}{r_{i}+1} \right \rceil \!-\! \left \lceil \!\frac{k - \sum _{i \in [s-1]} r_{i}\left \lceil \frac{n_i}{r_{i}+1} \!\right \rceil }{r_{s}} \right \rceil\!\!.\!\!
\end{equation}
It was  proved that an optimal  $r$-local LRC  can be shortened to obtain an optimal ML-LRC with respect to bound (\ref{eq_SingletonBoundMultipleLocalities}). For the case of two localities ($s=2$), \cite{zeh} gave an explicit algorithm that described the structure of the parity-check matrix for an optimal ML-LRC. To the best of our knowledge, direct and structured constructions of optimal ML-LRCs with respect to bound (\ref{eq_SingletonBoundMultipleLocalities}) have not yet been obtained except the above shortening technique.

Just like that the $(r,\delta)$-locality generalizes the $r$-local LRC, it is naturally to add similar features to the ML-LRCs. In this paper, we introduce the LRCs with all-symbol multiple $(r_{i}, \delta_{i})_{i\in [s]}$-localities ($s\geq 2$). Comparing with ML-LRCs, an LRC with multiple $(r_{i}, \delta_{i})_{i\in [s]}$-localities could not only locally recover a single failed node,  but also tolerate multiple nodes failures in other nodes among every  $n_{i}$ nodes. Moreover, the parameters satisfy $r_{1}\leq \dots\leq r_{s}$ and $\delta_{1}\geq \dots\geq\delta_{s}\geq 2$, which make the code more useful and attractive in some practical scenarios, e.g., when the distributed storage system employs such a multiple $(r_{i}, \delta_{i})_{i\in [s]}$-localities LRC, some hot data can be repaired quickly with the smaller locality $r_{i}$ while having a better erasure-tolerance with the larger $\delta_{i}$. By employing the regenerating-set technique of Wang and Zhang \cite{Wang1}, we derive a Singleton-like upper bound on the minimum distance for the LRCs with multiple $(r_{i}, \delta_i)_{i\in [s]}$-localities. Then, we construct a class of explicit and structured optimal ML-LRCs by employing the parity-splitting technique, and further extend them to the cases of multiple $(r_{i}, \delta)_{i\in [s]}$-localities.

The rest of this paper is organized as follows. In Section \ref{pre}, the concept of regenerating sets in \cite{Wang1} are recalled. In Section \ref{bound},  we firstly deal with the case of two $(r_{i}, \delta_{i})_{i\in\{1, 2\}}$ localities, and then the general one of multiple $(r_{i}, \delta_{i})_{i\in [s]}$ ($s\geq 2$) localities, where Singleton-like bounds are obtained. Section \ref{construction} studies the optimal constructions. Finally, we conclude the paper in section \ref{conclusion}.

\section{Preliminaries}\label{pre}
In this section, we give some preliminaries of regenerating sets, which was proposed by Wang and Zhang \cite{Wang1} to prove some minimum distance bounds, e.g., the Singleton-like bounds and the integer programming-based bound \cite{Wang:IT}. We will also use this technique to derive bounds in the next section.

Let $\ff_q$ be a finite field with size $q$, where $q$ is a prime power. An $[n,k,d]_q$ linear code $\mathcal{C}$ is a $q$-ary linear code with length $n$, dimension $k$ and minimum distance $d$.
\begin{definition}[\upshape \cite{Wang1}]\label{DefRegSet}
For an $[n,k,d]_q$ linear code $\mathcal{C}$, a \emph{regenerating set} of the $i$-th coordinate, $1\leq i\leq n$, is a subset $R\subseteq [n]\triangleq\{1,2,\ldots,n\}$ such that $i\in R$ and $\bf{\vec{g}_i}$ is an $\mathbb{F}_q$-linear combination of $\{\bf{\vec{g}_j}\}_{j \in R\backslash \{i\}}$, where $\bf{\vec{g}_i}$ denotes the $i$-th column vector of the generator matrix $\bf{G}$ of code $\mathcal{C}$.
The collection of all regenerating sets of the $i$-th coordinate is denoted by $\mathcal{R}_i$. Furthermore, a sequence of regenerating sets $R_1, R_2,\dots,R_m$, where $R_i\in\mathcal{R}_{l_i}$ and $l_i\in[n]$ for $1\leq i\leq m$, is said to have a {\it nontrivial union} if $l_j\notin\cup_{i=1}^{j-1}R_i$ for $1\leq j\leq m$.
\end{definition}

{\em Remark 1:}
\label{def1}
For the regenerating set $R_{i}$ of the $i$-th coordinate, it is called \emph{minimal} if there is no proper subset $R'\subset R_{i}\backslash \{i\}$ such that $\bf{\vec{g}_i}$ is an $\mathbb{F}_q$-linear combination of $\{\bf{\vec{g}_j}\}_{j \in R'}$. In the rest of this paper, without the loss of generality, we always assume that a regenerating set is minimal, and under this stricter definition, the set $R$ is a regenerating set of each of its elements.
Moreover, on one hand, the regenerating sets $R_1, R_2,\dots,R_m$ with a nontrivial union implies that $R_j\not\subseteq\cup_{i=1}^{j-1}R_i$ for $1\leq j\leq m$; and on the other hand, if $R_j\not\subseteq\cup_{i=1}^{j-1}R_i$ for $1\leq j\leq m$, it is clear that there exist $l_i\in R_i$, $i=1,\ldots, m$, such that $l_j\notin\cup_{i=1}^{j-1}R_i$ for $1\leq j\leq m$.

For a linear code $\mathcal{C}$, define the function \cite{Wang1}
\begin{eqnarray*}
\Phi(x)&=&\min\{|\cup_{i=1}^xR_i|: R_i\in\mathcal{R}_{l_i}\\
&& \mbox{~and~} R_1,\dots,R_x \mbox{ have a nontrivial union}\}.
\end{eqnarray*}
It is easy to see that $\Phi(x+1)\ge \Phi(x)+1$, which implies that $\Phi(x+1)-(x+1)\ge \Phi(x)-x$, or $\Phi(x)-x$ is increasing by $x$. The following theorem gives a general upper bound of the minimum distance $d$.

\begin{proposition}[\upshape \cite{Wang1}]\label{key thm}
For an $[n,k,d]_q$ linear code,
\begin{eqnarray*}
d&\leq& n-k+1-\rho,\; \\
&&\mbox{where }~\rho=\max\{x : \Phi(x)-x<k\}.\nonumber
\end{eqnarray*}
\end{proposition}

Next we give an alternative proof of Proposition \ref{key thm} by employing a  parity-check matrix approach \cite{Hao}.
By Definition \ref{DefRegSet} and Remark 1, it is clear that $R_{i}$ is a (minimal) regenerating set of the $i$-th coordinate if and only if there exists a codeword (or parity-check equation) $\mathbf{e}_i$ in the dual code $\cC^\perp$ such that ${\rm supp}(\mathbf{e}_i)=R_{i}$, where ${\rm supp}(\cdot)$ denotes the support of a vector (or the set of its non-zero coordinates).
Suppose that $k+x>\Phi(x)$, $R_1,\ldots, R_x$ have a non-trivial union and $\Phi(x)=|\cup_{i=1}^x R_i|$. Let $\mathbf{e}_1,\ldots, \mathbf{e}_x$ be their corresponding parity-check equations. Since $R_j\not\subseteq\cup_{i=1}^{j-1}R_i$ for $1\leq j\leq x$, it is clear that $\mathbf{e}_1,\ldots, \mathbf{e}_x$ are linearly independent, which implies that $x\le n-k$. Let $\mathbf{H}$ be an $(n-k)\times n$ parity-check matrix of $\cC$, where  $\mathbf{e}_1,\ldots, \mathbf{e}_x$ form its first $x$ rows. By deleting the first $x$ rows and the columns in $\cup_{i=1}^x R_i$ of $\mathbf{H}$, we obtain an $(n-k-x)\times (n-\Phi(x))$ submatrix $\mathbf{H}'$. Let $\cC'$ be the $[n',k',d']_q$ linear code with the parity-check matrix $\mathbf{H}'$. Clearly, $n'=n-\Phi(x)$, $k'\ge k+x-\Phi(x)>0$ and $d' \ge d$. Therefore, defining the largest possible minimum distance of an $[n',k']_q$ linear code by $d_q^{\rm(opt)}(n',k')$, we have the following result \cite{GHW}.

\begin{proposition} \label{cmt}
For an $[n,k,d]_q$ linear code,
\begin{eqnarray}
\label{eqcmt}
d&\leq& \min_{1\le x\le \rho} d_q^{\rm(opt)}(n-\Phi(x), k+x-\Phi(x)),\; \\
&&\mbox{where }~\rho=\max\{x : \Phi(x)-x<k\}.\nonumber
\end{eqnarray}
\end{proposition}

By invoking the well known Singleton bound for $x=\rho$ in the right-hand side of (\ref{eqcmt}), we have that $d\le n-k+1-\rho$. Hence, Proposition \ref{key thm} is a naturally corollary of Proposition \ref{cmt}.

Definition of the $(r, \delta)$-locality proposed in \cite{prakash2012optimal} could be redefined in regenerating-set language as follows.
\begin{definition}[\upshape \cite{Wang1}]\label{formal definition}
The $i$-th coordinate, $1 \leq i \leq n$, of an $[n, k, d]_q$ linear code $\cC$ is said to have $(r,\delta)$-locality if there exists a subset $S_{i}\subseteq [n]$ satisfying
\begin{itemize}
  \item[(1)~]$i\in S_{i} $, $\delta\leq |S_{i}|\leq r+\delta-1$; and
  \item[(2)~] For any $E\subseteq S_{i}$ with $|E|=\delta-1$, and for any $j\in E$, it has $(S_{i}-E)\cup\{j\}\in\mathcal{R}_{j}$.
\end{itemize}
\end{definition}
\smallskip
\section{Upper Bounds for Codes with Multiple    $(r_{i}, \delta_{i})_{i\in[s]}$-localities}\label{bound}

In this section, we firstly  define LRCs with two $(r_{i}, \delta_{i})_{i\in\{1, 2\}}$ localities and provide an upper bound on the minimum distance $d$ by employing the regenerating set technique. Then, we extend the bound to LRCs with $(r_{i}, \delta_{i})_{i\in[s]}$ ($s\geq2$)-localities similar to \cite{zeh}.
The definition of LRCs with two $(r_{i}, \delta_{i})_{i\{1,2\}}$ localities  follows.
\begin{definition}\label{new definition1}
Let $\mathcal{T}_{1}\subseteq[n]$ and $\mathcal{T}_{2}=[n]\diagdown \mathcal{T}_{1}$ be two distinct sets with $|\mathcal{T}_{i}|=n_{i}$ for $i=1, 2$. Let $r_1$, $r_2$, $\delta_1$, $\delta_2$ be integers such that $r_{1}\leq r_{2}$, $\delta_{1}\geq\delta_{2}\geq 2$. An $[n, k, d]_q$ linear code $\cC$ is said to have two $(r_{i},\delta_{i})_{i\in\{1,2\}}$ localities if for $i=1,2$ and each coordinate $\iota\in \mathcal{T}_{i}$, there exist a subset $S_{\iota}\subseteq \mathcal{T}_{i}$ satisfying
\begin{itemize}
  \item[(1)~]$\iota\in S_{\iota} $, $\delta_{i}\leq |S_{\iota}|\leq r_{i}+\delta_{i}-1$; and
  \item[(2)~] For any $E\subseteq S_{\iota}$ with $|E|=\delta_{i}-1$, and for any $j\in E$, it has $(S_{\iota}-E)\cup\{j\}\in\mathcal{R}_{j}$.
\end{itemize}
\end{definition}

\begin{lemma}\label{lem3.1}
For an $[n,k,d]$ LRC $\cC$ with two $(r_{i}, \delta_{i})_{i\in\{1,2\}}$ localities, if  $r_{1}\lceil n_{1}/(r_{1}+\delta_{1}-1)\rceil \leq k-1$ and $r_1\lceil(\Delta-1)/(\delta_{1}-1)\rceil+(\Delta-1)<n_1$. Then
\begin{small}
\begin{eqnarray*}&\label{OQPSK}
\Phi(x)\leq
\begin{cases}
r_{1} \left\lceil\frac{x}{\delta_{1}-1}\right\rceil+x, \quad\mbox{if $0\leq x\leq \Delta$};\\
r_{1}\left\lceil\frac{n_{1}}{r_{1}+\delta_{1}-1}\right\rceil+r_{2}\left\lceil\frac{x-\Delta}{\delta_{2}-1}\right\rceil+x,\quad\mbox{if $\Delta\le x\leq\rho+1$},
\end{cases}
\end{eqnarray*}
\end{small}
where $\Delta\triangleq\left\lceil n_{1}/(r_{1}+\delta_{1}-1)\right\rceil\big(\delta_{1}-1\big)$.
\end{lemma}
\begin{IEEEproof}
We prove the first part by induction on $x$ as Lemma 2 in \cite{Wang1}. It holds trivially for $x=0$. Assume it also holds  for $x\le x_{0}$, where $0\le x_{0}\le \Delta-1$. Denote $x_{0}+1=a(\delta_{1}-1)+b$, $a\in\mathbb{Z}$, $b\in[\delta_{1}-1]$. Let $T_{a(\delta_{1}-1)} = R_1 \cup \cdots \cup R_{a(\delta_{1}-1)}$ be a nontrivial union of $a(\delta_{1}-1)$ regenerating sets such that $\Phi(a(\delta_{1}-1)) = |T_{a(\delta_{1}-1)}|$. Note that $a(\delta_1-1)\le x_{0}\le\Delta-1$, thus
\begin{eqnarray*}
\Phi(a(\delta_{1}-1)) = |T_{a(\delta_{1}-1)}|\le r_1\left\lceil\frac{\Delta-1}{\delta_{1}-1}\right\rceil+(\Delta-1)<n_{1},
\end{eqnarray*}
which implies $\mathcal{T}_{1}\diagdown T_{a(\delta_{1}-1)} \neq\emptyset$. There are two cases:
\begin{itemize}
\item There exists $\iota_{0} \in\mathcal{T}_{1} \diagdown  T_{a(\delta_{1}-1)}$ such that $|S_{\iota_{0}}\diagdown  T_{a(\delta_1-1)}| \geq \delta_1-1\geqslant b$. Let $E \subseteq S_{\iota_{0}} \diagdown  T_{a(\delta_1-1)}$ with $|E| = \delta_1-1$.
Suppose $E = \{\iota_1,\cdots,\iota_{\delta_1-1}\}$.
Let $R_{\iota_j} = (S_{\iota_0} - E) \cup \{\iota_j\}$ for $j \in[\delta_1-1]$.
Then $R_{\iota_j}\in\mathcal{R}_{\iota_j}$ and $T_{a(\delta_1-1)}\cup (\cup_{j=1}^{b} R_{\iota_j})$ is a nontrivial union. We have
\begin{eqnarray*}
\Phi(x_0+1)& \le& |T_{a(\delta_1-1)} \cup R_{\iota_1} \cup \cdots \cup R_{\iota_b}| \\
& \le & \Phi(a(\delta_1-1)) + |S_{\iota_0} -E|+b \\
& \le & ar_1+a(\delta_1-1)+r_1+b\\
& = & r_1\left \lceil \frac{x_0+1}{\delta_1-1}\right \rceil +x_0+1.
\end{eqnarray*}
\item If for  any $\iota \in \mathcal{T}_{1} \diagdown T_{a(\delta_1-1)}$, $|S_\iota \diagdown  T_{a(\delta_1-1)}| < \delta_1-1$. Let $R_\iota= (S_\iota \cap T_{a(\delta_1-1)}) \cup \{\iota\}$, we have
$R_\iota \in \mathcal{R}_\iota$.
If $n_1 -|T_{a(\delta_1-1)}| \ge b$, we can choose $ \iota_1,\cdots, \iota_b \in\mathcal{T}_{1}\diagdown  T_{a(\delta_1-1)}$ such that $T_{a(\delta_1-1)}\cup (\cup_{j=1}^{b} R_{\iota_j})$ is a nontrivial union. Thus
\begin{eqnarray*}
\Phi(x_0+1) & \le & |T_{a(\delta_1-1)}\cup R_{ \iota_1} \cup \cdots \cup R_{ \iota_b}| \\
& = & |T_{a(\delta_1-1)}|+b \\
& = & \Phi(a(\delta_1-1)) +b \\
& \le & r_1\left \lceil \frac{x_0+1}{\delta_1-1}\right \rceil +x_0+1.
\end{eqnarray*}
If $n_1-|T_{a(\delta_1-1)}| < b$, since $r_1\lceil(\Delta-1)/(\delta_{1}-1)\rceil+(\Delta-1)<n_1$, then
$$\Phi(x_0+1)\leq n_1 < |T_{a(\delta_1-1)}|+b\le  r_1 \left\lceil \frac{x_0+1}{\delta_1-1}\right \rceil +x_0+1.$$
\end{itemize}

For the second part, we firstly prove that $\rho+1>\Delta$ or $\rho\geq\Delta.$ 	By the first part,
\begin{eqnarray*}
\Phi(\Delta)-\!\Delta
\leq \! r_{1}\left\lceil \frac{\Delta}{\delta_1-1}\right\rceil \!=r_1\left\lceil \frac{n_{1}}{r_{1}+\delta_{1}-1}\right\rceil\leq\! k-1\!<k.
\end{eqnarray*}
By the definition of $\rho$ in Proposition \ref{key thm}, we have $\rho\geq\Delta$. Therefore, we can similarly obtain the result by induction on $(x-\Delta)$ for $\Delta\le x\leq\rho+1$.

Note that if $y\triangleq x-\Delta=0$, the result holds trivially. Assume it also holds  for $y\le y_{0}$, where $0\le y_{0}\le \rho-\Delta$. Denote $y_{0}+1=a(\delta_{2}-1)+b$, $a\in\mathbb{Z}$, $b\in[\delta_{2}-1]$. Let $T_{\Delta+a(\delta_{2}-1)}$ be a nontrivial union of $\Delta+a(\delta_{2}-1)$ regenerating sets such that $\Phi(\Delta+a(\delta_{2}-1)) = |T_{\Delta+a(\delta_{2}-1)}|$. There are two cases:
\begin{itemize}
\item There exists $\iota_{0} \in [n]\diagdown T_{\Delta+a(\delta_{2}-1)}$ such that $|S_{\iota_{0}}\diagdown  T_{\Delta+a(\delta_{2}-1)}| \geq \delta_2-1\geqslant b$. Let $E \subseteq S_{\iota_{0}} \diagdown   T_{\Delta+a(\delta_{2}-1)}$ with $|E| = \delta_2-1$.
Suppose $E = \{\iota_1,\cdots,\iota_{\delta_2-1}\}$.
Let $R_{\iota_j} = (S_{\iota_0} - E) \cup \{\iota_j\}$ for $j \in[\delta_2-1]$.
Then $R_{\iota_j}\in\mathcal{R}_{\iota_j}$ and $T_{\Delta+a(\delta_{2}-1)}\cup (\cup_{j=1}^{b} R_{\iota_j})$ is a nontrivial union no matter $\iota_{0} \in\mathcal{T}_{1}$ or $\iota_{0}\in\mathcal{T}_{2}$. It follows that
\begin{eqnarray*}
\Phi(\Delta+y_0+1)& \le& |T_{\Delta+a(\delta_{2}-1)}\cup R_{\iota_1} \cup \cdots \cup R_{\iota_b}| \\
& \le &\Phi(\Delta+a(\delta_{2}-1)) + |S_{\iota_0} -E|+b \\
& \le & r_{1}\left\lceil\frac{n_{1}}{r_{1}+\delta_{1}-1}\right\rceil+ar_2\\
&&+\Delta+a(\delta_2-1)+r_2+b\\
& = &r_{1}\left\lceil\frac{n_{1}}{r_{1}+\delta_{1}-1}\right\rceil+r_2\left \lceil \frac{y_0+1}{\delta_2-1}\right \rceil\\
&& +\Delta+y_0+1.
\end{eqnarray*}
\item If for  any $\iota \in [n] \diagdown T_{\Delta+a(\delta_{2}-1)}$, $|S_\iota \diagdown T_{\Delta+a(\delta_{2}-1)}| < \delta_2-1\leqslant \delta_1-1$. Let $R_\iota= (S_\iota \cap T_{\Delta+a(\delta_{2}-1)}) \cup \{\iota\}$, we have
$R_\iota \in \mathcal{R}_\iota$.
If $n -|T_{\Delta+a(\delta_{2}-1)}| \ge b$, we can choose $ \iota_1,\cdots, \iota_b \in [n]\diagdown  T_{\Delta+a(\delta_{2}-1)}$ such that $T_{\Delta+a(\delta_{2}-1)}\cup (\cup_{j=1}^{b} R_{\iota_j})$ is a nontrivial union no matter $\iota_{0} \in\mathcal{T}_{1}$ or $\iota_{0}\in\mathcal{T}_{2}$.  Thus
\begin{eqnarray*}
\Phi(\Delta+y_0+1) & \le & |T_{\Delta+a(\delta_{2}-1)}\cup R_{ \iota_1} \cup \cdots \cup R_{ \iota_b}| \\
& = & |T_{\Delta+a(\delta_{2}-1)}|+b \\
& \le&r_{1}\left\lceil\frac{n_{1}}{r_{1}+\delta_{1}-1}\right\rceil+r_2\left \lceil \frac{y_0+1}{\delta_2-1}\right \rceil\\
&& +\Delta+y_0+1.
\end{eqnarray*}
If $n-|T_{\Delta+a(\delta_{2}-1)}| < b$, then
\begin{eqnarray*}
&&\Phi(\Delta+y_0+1) \leqslant n <|T_{\Delta+a(\delta_{2}-1)}| +b\\
&\leqslant&r_{1}\left\lceil\frac{n_{1}}{r_{1}+\delta_{1}-1}\right\rceil+r_2\left \lceil \frac{y_0+1}{\delta_2-1}\right \rceil +\Delta+y_0+1.
\end{eqnarray*}
\end{itemize}
\end{IEEEproof}

\begin{theorem}\label{newbound1}
For an $[n,k,d]$ LRC $\cC$ with two $(r_{i}, \delta_{i})_{i\in\{1,2\}}$ localities, if $r_{1}\lceil n_{1}/(r_{1}+\delta_{1}-1)\rceil \leq k-1$ and $r_1\lceil(\Delta-1)/(\delta_{1}-1)\rceil+(\Delta-1)<n_1$. Then
\begin{eqnarray}\label{bound1}
d \!\!&\le&\!\! n - k+1 -\left\lceil n_{1}/(r_{1}+\delta_{1}-1)\right\rceil\big(\delta_{1}-1\big)\nonumber\\
&&\!\!\!\!\!-\left(\left\lceil\frac{k-r_{1}\lceil n_{1}/(r_{1}+\delta_{1}-1)\rceil}{r_{2}}\right\rceil-1\right) \big(\delta_{2}-1\big).
\end{eqnarray}
\end{theorem}
\begin{IEEEproof}
By the definition of $\rho$ and Lemma \ref{lem3.1}, \begin{eqnarray*}
k&\leq&\Phi(\rho+1)-(\rho+1)\\
&\leq& r_{1}\left\lceil\frac{n_{1}}{r_{1}+\delta_{1}-1}\right\rceil+r_{2}\left\lceil\frac{(\rho+1)-\Delta}{\delta_{2}-1}\right\rceil.
\end{eqnarray*}
It follows that $$\left\lceil\frac{k-r_{1}\lceil n_{1}/(r_{1}+\delta_{1}-1)\rceil}{r_{2}}\right\rceil\leq \left\lceil\frac{(\rho+1)-\Delta}{\delta_{2}-1}\right\rceil,$$
thus
\begin{eqnarray*}
&&\!\!\!\!\!\!\!\!\!\!\!\!\!\!\!\!\!\!\!\!\left(\left\lceil\frac{k-r_{1}\lceil n_{1}/(r_{1}+\delta_{1}-1)\rceil}{r_{2}}\right\rceil-1\right)(\delta_{2}-1)
\\
&\leq& \left(\left\lceil\frac{(\rho-\Delta)+1}{\delta_{2}-1}\right\rceil-1\right)(\delta_{2}-1)
\leq \rho-\Delta,
\end{eqnarray*}
or
$$\rho\;\geq\;\Delta+\left(\left\lceil\frac{k-r_{1}\lceil n_{1}/(r_{1}+\delta_{1}-1)\rceil}{r_{2}}\right\rceil-1\right)(\delta_{2}-1).$$
Hence, we have the desired bound (\ref{bound1}) by Proposition \ref{key thm}.
\end{IEEEproof}

\medskip
{\em Remark 2:}
For $\delta_{1}=\delta_{2}=2$, the condition $r_1\lceil(\Delta-1)/(\delta_{1}-1)\rceil+(\Delta-1)<n_1$ is naturally satisfied, and the bound (\ref{bound1}) reduces to the bound (2) in \cite{zeh}. For $\delta_{1}>2$,  the condition $r_1\lceil(\Delta-1)/(\delta_{1}-1)\rceil+(\Delta-1)<n_1$, i.e., $r_1+\delta_1-1\mid n_1$.  Note that an LRC with two $(r_{i}, \delta_{i})_{i\in\{1,2\}}$ localities is also an $(r_{2}, \delta_{2})$-locality LRC, it is easy to verify that the bound (\ref{bound1}) is usually tighter than the bound (\ref{eq_GeneralizedSingleton}) for $r=r_2$ and $\delta=\delta_2$. If the condition of Theorem \ref{newbound1} is not satisfied, or $r_{1}\lceil n_{1}/(r_{1}+\delta_{1}-1)\rceil \geq k>k-1$,  then $$\left\lfloor(k-1)/r_{1}\right\rfloor(\delta_{1}-1)<\left\lceil n_{1}/(r_{1}+\delta_{1}-1)\right\rceil\big(\delta_{1}-1\big)=\Delta.$$
Hence, by Lemma \ref{lem3.1}, we have
\begin{eqnarray*}
&&\!\!\Phi\left(\left\lfloor(k-1)/r_{1}\right\rfloor(\delta_{1}-1)\right)-\left\lfloor (k-1)/r_{1}\right\rfloor(\delta_{1}-1)\\
&&\leq\; r_{1}\left\lfloor(k-1)/r_{1}\right\rfloor\leq k-1<k.
\end{eqnarray*}
By the definition of $\rho$, we obtain $\rho\geq\left\lfloor(k-1)/r_{1}\right\rfloor(\delta_{1}-1)=\left(\left\lceil k/r_{1}\right\rceil-1\right)(\delta_{1}-1)$. Therefore, by Proposition \ref{key thm}, we have
\begin{equation}\label{eqa1}
d\leq n-k+1-\left(\left\lceil k/ r_{1}\right\rceil-1\right)(\delta_{1}-1),
\end{equation}
which corresponds to the bound (\ref{eq_GeneralizedSingleton}) for a code with $(r_{1}, \delta_{1})$-locality. Note that if  $r_{1}\lceil n_{1}/(r_{1}+\delta_{1}-1)\rceil =k-1$, the bound (\ref{eqa1}) is identical with the bound (\ref{bound1}).

Definition \ref{new definition1} can be easily generalized to a code with  $(r_{i}, \delta_{i})_{i\in[s]}$ ($s\geq2$) localities, and the Singleton-like bound (\ref{bound1}) can also be generalized as follows.
\begin{definition}\label{new definition2}
Let $\mathcal{T}_{1},\mathcal{T}_{2},\ldots,\mathcal{T}_{s}$ be a partition of $[n]$, where $s\geq2$ and $|\mathcal{T}_{i}|=n_i, i\in[s]$. Let $r_1$, $r_2$, $\cdots$, $r_s$ and $\delta_1$, $\delta_2$, $\cdots$, $\delta_s$ be integers such that $r_{1}\leq r_{2}\leq\dots\leq r_{s}$, $\delta_{1}\geq\delta_{2}\geq\dots\geq\delta_{s}\geq 2$.  An $[n, k, d]_q$ linear code $\cC$ is said to have  multiple $(r_{i},\delta_{i})_{i\in[s]}$-localities if for $i=1,2,\ldots,s$ and each coordinate $\iota\in \mathcal{T}_{i}$, there exist a subset $S_{\iota}\subseteq \mathcal{T}_{i}$ satisfying
\begin{itemize}
  \item[(1)~]$\iota\in S_{\iota} $, $\delta_{i}\leq |S_{\iota}|\leq r_{i}+\delta_{i}-1$; and
  \item[(2)~] For any $E\subseteq S_{\iota}$ with $|E|=\delta_{i}-1$, and for any $j\in E$, it has $(S_{\iota}-E)\cup\{j\}\in\mathcal{R}_{j}$.
\end{itemize}
\end{definition}

\begin{lemma}\label{lem3.2}
For an  $[n,k,d]$ LRC $\cC$ with multiple $(r_{i}, \delta_{i})_{i\in[s]}$ ($s\geq2$) localities, if
$\;\sum_{i=1}^{s-1}r_{i}\lceil n_{i}/(r_{i}+\delta_{i}-1)\rceil \leq k-1$ and
\begin{eqnarray}\label{condition}
r_j\left\lceil\frac{\Delta_j-\Delta_{(j-1)}-1}{\delta_j-1}\right\rceil+\left(\Delta_j-\Delta_{(j-1)}-1\right)<n_j,\end{eqnarray}
where $\Delta_{0}\triangleq0$ and $\Delta_{j}\!\!\triangleq\sum_{i=1}^{j}\lceil n_{i}/(r_{i}+\delta_{i}-1)\rceil(\delta_{i}-1)$,
$j=1,2,\ldots,s-1$.
 Then
\begin{itemize}
\item For $\;\Delta_{(j-1)}\leq x\leq \Delta_{j}$, $j=1,2,\ldots, s-1$,
\begin{eqnarray*}
\Phi(x)\leq\! \sum_{i=1}^{j-1}r_{i}\left\lceil\frac{n_{i}}{r_{i}+\delta_{i}-1}\right\rceil+r_{j}\left\lceil\frac{x-\Delta_{(j-1)}}{\delta_{j}-1}\right\rceil+x;
\end{eqnarray*}

\item For $\;\Delta_{(s-1)}\leq x\leq\rho+1$,
\begin{eqnarray*}
\Phi(x)\leq\!\sum_{i=1}^{s-1}r_{i}\left\lceil\frac{n_{i}}{r_{i}+\delta_{i}-1}\right\rceil+r_{s}\left\lceil\frac{x-\Delta_{(s-1)}}{\delta_{s}-1}\right\rceil+x,
\end{eqnarray*}
\end{itemize}
\end{lemma}
\begin{IEEEproof}
For $\Delta_{(j-1)}\leq x\leq \Delta_{j}$, $j\in[s-1]$, we can easily prove the results by employing the method of induction as Lemma \ref{lem3.1};
For $\;\Delta_{(s-1)}\leq x\leq\rho+1$;
note that the condition $\sum_{i\in[s-1] }r_{1}\lceil n_{i}/(r_{i}+\delta_{i}-1)\rceil \leq k-1$ also ensures that $\rho+1>\Delta_{(s-1)}$.
Therefore, we can similarly obtain the result by induction on $(x-\Delta_{(s-1)})$ as Lemma \ref{lem3.1}.
\end{IEEEproof}

\begin{theorem}\label{newbound2}
For an $[n,k,d]$ LRC $\cC$ with multiple $(r_{i}, \delta_{i})_{i\in[s]}$ localities ($s\geq2$) satisfying the condition stated in Lemma \ref{lem3.2}, then
\begin{eqnarray}\label{bound2}
d \!\!&\le&\!\! n - k+1 -\sum_{i=1}^{s-1}\left\lceil\frac{n_{i}}{r_{i}+\delta_{i}-1}\right\rceil\big(\delta_{i}-1\big)\nonumber\\
&&\!\!\!\!\!-\left(\left\lceil\frac{k-\sum_{i=1}^{s-1}r_{i}\lceil \frac{n_{i}}{r_{i}+\delta_{i}-1}\rceil}{r_{s}}\right\rceil-1\right)\big(\delta_{s}-1\big).
\end{eqnarray}
\end{theorem}
\begin{IEEEproof}
By the definition of $\rho$ and Lemma \ref{lem3.2},
\begin{eqnarray*}
k&\leq&\Phi(\rho+1)-(\rho+1)\\
&\leq& \sum_{i=1}^{s-1}r_i\left\lceil\frac{n_{i}}{r_{i}+\delta_{i}-1}\right\rceil+r_{s}\left\lceil\frac{(\rho+1)-\Delta_{(s-1)}}{\delta_{s}-1}\right\rceil.
\end{eqnarray*}
By a little calculation, we have that
$$\rho\;\geq\; \Delta_{(s-1)}+\left(\left\lceil\frac{k-\sum_{i=1}^{s-1}r_{i}\lceil \frac{n_{i}}{r_{i}+\delta_{i}-1}\rceil}{r_{s}}\right\rceil-1\right)(\delta_{s}-1),$$
which implies the bound (\ref{bound2}) by Proposition \ref{key thm}.
\end{IEEEproof}
{\em Remark 3:}
For $\delta_{1}=\delta_{2}=\dots=\delta_{s-1}=\delta_{s}=2$, the condition (\ref{condition}) is naturally satisfied, and the bound (\ref{bound2}) reduces to the bound (3) in \cite{zeh}. Otherwise, for $\delta_{i_{0}}>2$, where $i_{0}=\mbox{max}\{i\in[s-1]\mid \delta_{i}>2\}$, the condition (\ref{condition}) , i.e., $r_i+\delta_i-1\mid n_i$, $i=1, \dots, i_{0}$. Moreover, the bound (\ref{bound2}) is also tighter than  the $(r_{s},\delta_{s})$-bound (\ref{eq_GeneralizedSingleton}).

\begin{corollary}\label{cor1}
Assume the condition in Theorem \ref{newbound2} is satisfied  and $\delta_{1}=\delta_{2}=\dots=\delta_{s}=\delta\geq 2$, then bound (\ref{bound2}) reduces to
\begin{eqnarray}\label{bound3}
\small
 d \le n - k+1-\left(\Gamma-1\right)\big(\delta-1\big),
\end{eqnarray}
where $\Gamma\!=\sum_{i=1}^{s-1}\left\lceil \frac{n_{i}}{r_{i}+\delta-1}\right\rceil+\left\lceil\frac{k-\sum_{i=1}^{s-1}r_{i}\lceil  n_{i}/(r_{i}+\delta-1)\rceil}{r_{s}}\right\rceil$.
\end{corollary}

\smallskip
\section{Optimal Constructions of  ML-LRCs and LRCs with Multiple $(r_{i}, \delta)_{i\in [s]}$-localities}\label{construction}

In this section, based on the parity-splitting technique of Reed-Solomon (RS) codes in \cite{prakash2012optimal}, we firstly give an explicit and structured optimal ML-LRCs meeting the bound (\ref{eq_SingletonBoundMultipleLocalities}). Then the proposed constructions are generalized to the LRCs with multiple $(r_{i}, \delta)_{i\in [s]}$-localities, which are optimal with respect to the bound (\ref{bound3}).

\begin{theorem}\label{con1}
Let $n=\sum_{i\in [s]}n_{i}$, $s\geq2$. Let $r_1$, $r_2$, $\cdots$, $r_s$ be integers such  that  $2\leq r_{1}\leq r_{2}\leq\dots\leq r_{s}$, $r_{i}+1\mid n_{i}$, $i\in[s]$ and $\frac{n_{s}}{r_{s}+1}=\left\lceil\frac{k-\sum_{i\in[s-1]}r_i(n_{i}/(r_{i}+1))}{r_{s}}\right\rceil$. If $q > n$, then there exists an explicit and optimal ML-LRC over $\mathbb{F}_q$.
\end{theorem}
\begin{IEEEproof}
 Let $H'$ be the parity check matrix of an $[n,k',d']$ RS code over $\mathbb{F}_q$, where
 \begin{eqnarray*}
k'&=& k+ \sum_{i\in[s-1]}\frac{n_{i}}{r_{i}+1}+\left\lceil\frac{k-\sum_{i\in[s-1]}r_i\frac{n_{i}}{r_{i}+1}}{r_{s}}\right\rceil-1\\
&=&k+\sum_{i\in[s]}\frac{n_{i}}{r_{i}+1}-1,\\
d' &=& n-k'+1 \\
&=&\!\!\!\! n-k+2- \!\!\!\! \sum_{i\in[s-1]}\frac{n_{i}}{r_{i}+1}-\left\lceil\frac{k-\sum_{i\in[s-1]}r_i\frac{n_{i}}{r_{i}+1}}{r_{s}}\right\rceil.
\end{eqnarray*}
  Note that  if $q > n$, such RS code always exists, and
$H'_{(n-k') \times n}$can be equivalently transformed into a Vandermonde  matrix, i.e.,
\begin{equation}
H'= \left[ \begin{array}{c}
    {1,1,\dots,1}_{1 \times n} \\
     A_{(n-k'-1)\times n}
    \end{array} \right].
\end{equation}

Consider the code
$\mathcal{C}$ whose parity check matrix $H$ obtained by splitting the first  row of $H'$ as follows:
\begin{equation}
 H=\left[ \begin{array}{ccc}
            A_1 &&\\
	    & \ddots &\\
	    &&A_{s}\\
	    \hline
	    & A _{(n-k'-1)\times n}&
           \end{array}
\right],
\end{equation}
\begin{equation*}
 \mbox{where}\quad A_{i}\;=\;\left[ \begin{array}{ccc}
            \underbrace{11\dots1}_{r_{i}+1}&&\\
	    & \ddots &\\
	    &&   \underbrace{11\dots1}_{r_{i}+1}
           \end{array}
\right]_{\frac{n_{i}}{r_{i}+1}\times n_{i}},\; i\in[s].
\end{equation*}

Firstly, we have $\text{dim}(\mathcal{C})\ge k$, which is due to the structure of $H$ and
$$\text{dim}(\mathcal{C}^{\bot})=n-\text{dim}(\mathcal{C})\leq n-k'-1+\sum_{i\in [s]}\frac{n_{i}}{r_{i}+1}=n-k.$$
Then we have
\begin{eqnarray}\label{ineq1}
d\!\!\!\!&\overset{(a)}{\geqslant}&d'
=n-k+2\nonumber\\
&&-\sum_{i\in[s-1]}\frac{n_{i}}{r_{i}+1}-\!\!\left\lceil\frac{k-\!\!\!\sum_{i\in[s-1]}r_i\frac{n_{i}}{r_{i}+1}}{r_{s}}\right\rceil
\end{eqnarray}
and
\begin{eqnarray}\label{ineq2}
d\!\!&\overset{(b)}{\leqslant}&\!\!n-\text{dim}(\mathcal{C})+2\nonumber\\
&-&\!\!\!\!\!\!\sum_{i\in[s-1]}\frac{n_{i}}{r_{i}+1}-\left\lceil\frac{\text{dim}(\mathcal{C})-\!\!\sum_{i\in[s-1]}r_i\frac{n_{i}}{r_{i}+1}}{r_{s}}\right\rceil,
\end{eqnarray}
where $(a)$ follows from the fact that $\cC$ is a subcode of the RS code, $(b)$ follows because the structure of $H$ ensures the code $\cC$ has multiple localities and the condition $\frac{n_{s}}{r_{s}+1}=\left\lceil\frac{k-\sum_{i\in[s-1]}r_i(n_{i}/(r_{i}+1))}{r_{s}}\right\rceil$ implies $\sum_{i\in[s-1]}r_i n_{i}/(r_{i}+1)\leq k-1$.
Combining (\ref{ineq1}) and (\ref{ineq2}), we have $\text{dim}(\mathcal{C})\le k$.

Thus $\text{dim}(\mathcal{C})=k$.
Combining (\ref{ineq1}) and (\ref{ineq2}), the optimality of $\cC$ can also be obtained.
\end{IEEEproof}

Based on the construction above, we can easily obtain the following general construction.

\begin{theorem}\label{con2}
Let $n=\sum_{i\in s}n_{i}$, $s\geq2$. Let $r_1$, $r_2$, $\cdots$, $r_s$, $\delta$ be integers such  that  $2\leq r_{1}\leq r_{2}\leq\dots\leq r_{s}$, $r_{i}+\delta-1\mid n_{i}$, $i\in[s]$ and $\frac{n_{s}}{r_{s}+\delta-1}=\left\lceil\frac{k-\sum_{i\in[s-1]}r_i(n_{i}/(r_{i}+\delta-1))}{r_{s}}\right\rceil$. If $q > n$, then there exists an explicit and optimal multiple $(r_{i}, \delta)_{i\in[s]}$-localities LRC with respect to bound (\ref{bound3}).
\end{theorem}
\begin{IEEEproof}
Similar to Theorem $8$ in \cite{prakash2012optimal}, let $H'$ be the parity check matrix of an $[n,k',d']$ RS code over $\mathbb{F}_q$, where
$$
k'= k+(\Gamma-1)(\delta-1)=k+\left(\sum_{i\in[s]}\frac{n_{i}}{r_{i}+\delta-1}-1\right)(\delta-1),
$$
and $\Gamma\!\triangleq\sum_{i\in[s-1]} \frac{n_{i}}{r_{i}+\delta-1}+\left\lceil\frac{k-\sum_{i\in[s-1]}r_{i}(n_{i}/(r_{i}+\delta-1))}{r_{s}}\right\rceil$. Thus
$$d' = n-k'+1  = n-k+1- (\Gamma-1)(\delta-1).$$ We also take $H'$ to be a Vandermonde matrix as follow:
\begin{equation}
H'= \left[ \begin{array}{c}
     Q_{(\delta-1) \times n} \\
     A_{(n-k'+1-\delta)\times n}
    \end{array} \right].
\end{equation}
 We  partition the matrix $Q$ in terms of submatrices as:
\begin{equation*}
Q = \left [ Q^{(1)}_{1} Q^{(1)}_{2} \ldots Q^{(1)}_{\frac{n_{1}}{r_{1}+{\delta}-1}} \mid \ldots \mid Q^{(s)}_{1} Q^{(s)}_{2} \ldots Q^{(s)}_{\frac{n_{s}}{r_{s}+{\delta}-1}}  \right],
\end{equation*}
where
$Q^{(i)}_{j}, i\in[s], j\in[n_{i}/(r_{i}+\delta-1)]$  are matrices of size $ (\delta-1) \times (r_{i}+\delta - 1)$.

By employing the parity-splitting technique, we consider the code $\mathcal{C}$ whose parity check matrix $H$ as follow:
\begin{equation}
 H=\left[ \begin{array}{ccc}
            A_1 &&\\
	    & \ddots &\\
	    &&A_{s}\\
	    \hline
	    & A _{(n-k'+1-\delta)\times n}&
           \end{array}
\right],
\end{equation}
where
\begin{equation*}
 A_{i}\;=\;\left[ \begin{array}{ccc}
           Q^{(i)}_{1} &&\\
	    & \ddots &\\
	    &&  Q^{(i)}_{\frac{n_{i}}{r_{i}+{\delta}-1}}
           \end{array}
\right]_{\frac{(\delta-1)n_{i}}{r_{i}+\delta-1}\times n_{i}}, \;i\in[s].
\end{equation*}

Clearly, the structure of $H$ ensures the code $\cC$ has multiple $(r_{i}, \delta)_{i\in[s]}$-localities and $\cC$ is a subcode of the RS code, it is not hard to prove that $\text{dim}(\mathcal{C})=k$ and the optimality of $d$ with respect to the bound (\ref{bound3}) similar  to Theorem \ref{con1}.
\end{IEEEproof}
\smallskip
\section{Conclusion}\label{conclusion}
We introduced LRCs  with multiple $(r_{i}, \delta_{i})_{i\in [s]}$-localities, which is useful and attractive in some practical scenarios, especially for the hot data in distributed storage systems. An upper bound on the minimum distance was obtained for LRCs  with multiple $(r_{i}, \delta_{i})_{i\in [s]}$-localities, which extended the bound of ML-LRCs in \cite{zeh}. By employing the parity-splitting technique of \cite{prakash2012optimal}, we gave optimal constructions with $r_{1}\leq r_{2}\leq\dots\leq r_{s}$ and $\delta_{1}=\delta_{2}=\dots=\delta_{i}=\delta$, $\delta\geq 2$. Future works might focus on constructing more explicit and structured optimal LRCs with multiple $(r_{i}, \delta_{i})_{i\in [s]}$-localities, especially for the parameters  $r_{1}<r_{2}<\dots<r_{s}$, $\delta_{1}>\delta_{2}>\dots>\delta_{s}$.





%

\end{document}